\documentclass[12pt]{article}
\usepackage{epsfig,equations,graphics,amssymb}
\usepackage{cite}

\begin{document}

\def\nicefrac#1#2{\hbox{${#1\over #2}$}}
\textwidth 10cm
\setlength{\textwidth}{13.7cm}
\setlength{\textheight}{23cm}

\oddsidemargin 1cm
\evensidemargin -0.2cm
\addtolength{\topmargin}{-2.5 cm}

\newcommand{\nn}{\nonumber}
\newcommand{\raw}{\rightarrow}
\newcommand{\be}{\begin{equation}}
\newcommand{\ee}{\end{equation}}
\newcommand{\bea}{\begin{eqnarray}}
\newcommand{\eea}{\end{eqnarray}}
\newcommand{\dl}{\stackrel{\leftarrow}{D}}
\newcommand{\dr}{\stackrel{\rightarrow}{D}}
\newcommand{\dd}{\displaystyle}
\newcommand{\Ln}{{\rm Ln}}

\pagestyle{empty}

\begin{flushright}
FTUAM 01/12 \\
IFT-UAM/CSIC 01/18  \\
hep--ph/0107147 \\
July 2001 \\
\end{flushright}
\vskip1cm

\renewcommand{\thefootnote}{\fnsymbol{footnote}}
\begin{center}
{\Large\bf
Effective Higgs-quark-quark couplings from a heavy supersymmetric spectrum}\\[1cm]
{\large Antonio Dobado$^a$,
Mar\'{\i}a J. Herrero$^b$
and
David Temes$^b$~\footnote{electronic addresses:
malcon@fis.ucm.es, herrero@delta.ft.uam.es, temes@delta.ft.uam.es}
}\\[6pt]
{\it $^a$ Departamento de F\'{\i}sica Te\'{o}rica \\
   Universidad Complutense de Madrid, 28040 Madrid, Spain. \\
$^b$ Departamento de F\'{\i}sica Te\'{o}rica \\
   Universidad Aut\'{o}noma de Madrid,
   Cantoblanco, 28049 Madrid, Spain.}
\\[1cm]

\begin{abstract}
\vspace*{0.4cm}

In this paper we study the Yukawa Higgs-quark-quark interactions that
are generated from radiative corrections of squarks and gluinos, in
the minimal supersymmetric standard model. We compute the corrections
to the effective action for Higgs and quark fields that are produced
by explicit integration in the path integral formalism of all the
squarks and gluinos at the one-loop level and order $\alpha_s$. In
addition, we consider the limit of nearly degenerate heavy squarks and
gluinos, with masses much 
larger than the electroweak scale, and we derive the
effective Lagrangian containing all the relevant new local
Higgs-quark-quark interactions. We show that these new interactions do
remain nonvanishing, even in the case of infinitely heavy
supersymmetric particles and, therefore, we demonstrate explicitly
the nondecoupling behavior of squarks and gluinos in Higgs bosons
physics. We present the set of new Yukawa couplings and finally derive
the corresponding one-loop, order $\alpha_s$, corrections to the Higgs
bosons partial decay widths into quarks.
\end{abstract}

\end{center}

\vfill
\clearpage

\renewcommand{\thefootnote}{\arabic{footnote}}
\setcounter{footnote}{0}

\pagestyle{plain}

\section{Introduction}

The Higgs particle discovery and the subsequent study of its physical
parameters and couplings is probably the most urgent challenge for
particle physics phenomenology in the near future. The major task of
this analysis, once one or several Higgs particles have been discovered, will be to
unravel their supersymmetric (SUSY) or nonsupersymmetric origin. Obviously,
the best evidence for the existence of supersymmetry in nature will
be the discovery  of genuine
SUSY particles, but it may well happen that the SUSY spectrum turns out
to be too heavy as to be produced directly in the planned
experiments. In this case, and in order to prove the existence of SUSY
particles, it will be mandatory to search for indirect SUSY signals via their
contribution to the radiative corrections in either couplings, physical
parameters, as particle masses and/or decay widths, or precision observables,
in a similar way as it
has been done in the past at CERN $e^+ e^-$ collider LEP for indirect searches of the top quark and
the
standard model (SM) Higgs particle. In fact, there is already an extensive work
done on these SUSY radiative corrections and their phenomenological
implications in the literature~\cite{proceedings}.

In this work we study the radiative corrections from heavy
SUSY particles to the Yukawa Higgs-quark-quark interactions using a
different but powerful approach provided by the effective
action formalism. We concentrate here on the contributions from
the SUSY-QCD sector of
the minimal supersymmetric standard model (MSSM) but extensions of
this work to the SUSY-electroweak sector of the MSSM and to other SUSY models
can be easily performed. In short, the procedure goes as
follows. We start with the tree-level MSSM
Lagrangian~\cite{MSSM,2HDM}, 
and consider
the Higgs sector interactions with quarks and the relevant SUSY-QCD
sector interactions. These are the Higgs-squark-squark interactions
and the squark-quark-gluino interactions. We perform a standard
functional integration, to one-loop level, of all the squarks and
gluinos and get the effective action containing all the exact,
non-local, one-loop level interactions of the MSSM Higgs particles with
quarks. We next examine the limit where the SUSY particles are much
heavier than the electroweak scale, $m_{EW} \sim \mathcal{O}(246 \, \, {\rm
GeV})$, and find out the
effective local interactions of these Higgs particles with quarks being
generated at one-loop level and at low energies from the heavy
SUSY-QCD spectrum. These effective local interactions come, in our
approach, from a large SUSY mass
expansion of the one-particle irreducible functions 
in powers of $(p/M_{SUSY})^n$,
with $p$ being any typical external
momentum or mass of the order of the electroweak scale, and $M_{SUSY}$
being the heavy SUSY masses ($M_{SUSY} \gg
m_{EW}$). The leading
contribution of zero order, $n=0$,
corresponds to the exact result in the zero external momentum
approximation. The advantage of using an effective action approach is
that it provides not just the first term in the expansion but the
whole series.

The main motivation of this work is to investigate with complete
generality whether there is or not decoupling of heavy squarks and
gluinos in the low energy effective interactions of the MSSM Higgs
sector with the SM quarks. In particular we will concentrate on the
effective Yukawa-like interactions $h^o b \bar b$,
$h^o t \bar t$, $H^o b \bar b$, $H^o t \bar t$, $A^o b \bar b$,
$A^o t \bar t$ and $H^+ \bar t b$ that are generated by
explicit integration of stops, sbottoms and gluinos to one-loop
level and to order $\alpha_S$. We have focused on these SUSY-QCD third
generation radiative corrections since they are the
dominant ones~\cite{Dabelstein,topQCD,solaHTB,cjs,EberlHTB}.

The decoupling behavior of SUSY particles has been studied
previously with complete generality for the effective electroweak
gauge bosons self-interactions and within the same effective action
approach in a series of works~\cite{TesisS}. It has been shown, by explicit
integration to one-loop level, that heavy
squarks, sleptons, charginos and neutralinos do decouple in the low
energy $W^{\pm}$, $Z$ and $\gamma$ self-interactions. In agreement
with the Appelquist-Carazzone theorem~\cite{Apple}, this means that
the heavy SUSY particles do not generate any new electroweak bosons
interactions that were not already present at the SM
Lagrangian, and all the terms found in the low energy expansion of the
effective action indeed decouple as inverse powers of the heavy SUSY masses.
The behavior of decoupling in the present context of
Higgs-quark-quark interactions has been studied partially by several
authors but none with complete generality. Some of these couplings
were studied in the context of the zero external approximation 
in~\cite{cmwpheno,kolda,CarenaDavid,EberlTodos}. The radiatively
generated momentum-dependent Yukawa couplings for the case of neutral
Higgs bosons and light quarks with $m_q \ll m_H$, and for external
on-shell particles, have been considered in~\cite{polonsky}.
In addition, there are several studies of the
partial decay widths $\Gamma (h^o \rightarrow b \bar
b)$~\cite{HaberTemes} and 
$\Gamma (H^+ \rightarrow t \bar b)$~\cite{ourHtb} in the limit of
heavy SUSY particles, by
using a Feynman diagrammatic approach. All these studies already
indicate a nondecoupling behavior of the heavy SUSY particles in the
Higgs sector of the MSSM. This is
manifested as some non-vanishing contributions in the one-loop radiative
corrections of the partial widths from genuine SUSY particles, which
are present even in the case of infinitely heavy SUSY masses.

Here we study the decoupling behavior of SUSY particles in the
effective field theory formalism. Indeed, we deduce the particular 
effective theory that remains at low energies after explicit
integration in the path integral of the 
heavy SUSY modes in the
MSSM theory. Our starting point with regard to the Higgs sector interactions
with quarks is the well known tree-level
MSSM Lagrangian, which belongs to the class of two Higgs doublet models
of type II (2HDMII)~\cite{2HDM} where one of the two doublets couples
just to the
top-like quarks and the other one just to the bottom-like quarks. At
this level, the absence of other possible Yukawa couplings is indeed a consequence of
the underlying supersymmetry.  When integrating out SUSY particles,
one could, in principle, find out different scenarios at low
energies. One may end up either with effective interactions of the same
type II or, in the most general case, one could generate new
interactions belonging to the unrestricted type III models 
(2HDMIII)~\cite{2HDMIII}, where both
doublets couple to both top and bottom-like quarks. It has been argued
on general grounds~\cite{HaberTemes,ourHtb} that the most plausible
scenario at low energies is a 2HDMIII, since the supersymmetry is broken by
the heavy SUSY masses and, therefore, the restrictions imposed by this
symmetry do not apply. However, the final answer can only be known once these
effective low energy interactions have been explicitly computed. In
case one gets a 2HDMII one should conclude that there is indeed decoupling
of the SUSY particles, since all their effects can be absorbed into
redefinitions of the low energy parameters, couplings and fields. In contrast, if one gets
a 2HDMIII one should conclude there is no decoupling of the
heavy SUSY particles, since their effects cannot be eliminated by
that redefinition. In this latter case, the new effective
interactions do remain at low energies, even in the
infinitely heavy SUSY masses limit and consequently, it may bring some
hint on indirect SUSY signals.

The final goal of our work is precisely to compute, with complete
generality, all these new effective Yukawa interactions in order to
provide, first, a formal proof of SUSY-non-decoupling behavior in
Higgs physics and, second, the
complete list with the particular values of the new $h^o b \bar b$,
$h^o t \bar t$, $H^o b \bar b$, $H^o t \bar t$, $A^o b \bar b$,
$A^o t \bar t$ and   $H^+ \bar t b$ Yukawa couplings,
which in turn have interesting applications for phenomenology
~\cite{cmwpheno,kolda,CarenaDavid,polonsky,EberlTodos,HaberTemes,ourHtb,RADCOR2000,HaberLC,MJsitges,siannah,curiel}.

The paper is organized as follows. In Section 2 the Higgs and SUSY
sectors together with the relevant interactions for the present work
are shortly reviewed. In Section 3 we integrate out the squarks and
gluino fields to one-loop level and $\mathcal{O}(\alpha_s)$
and we give the expressions for the non-local effective
action. In Section 4 we compute the effective local interactions of
Higgs particles with quarks in the limit of heavy squarks and
gluinos. Finally, Section 5 is devoted to a complete discussion of the
results and to the conclusions.   

\section{Higgs, SUSY-QCD sectors and relevant interactions in the MSSM}

In this section we shortly review the relevant MSSM spectrum and
interactions for the present work which are the Higgs sector, the
SUSY-QCD sector, the Yukawa
Higgs-quark-quark interactions, the squark-quark-gluino interactions
and the Higgs-squark-squark interactions.

In the MSSM Higgs sector there are five physical Higgs
particles: two $CP$-even neutral scalar particles
$h^{o}$, $H^{o}$, one $CP$-odd neutral pseudoscalar particle $A^{o}$ and two
charged scalar particles $H^{\pm}$. Due to supersy\-mme\-try, the parameters of
the Higgs sector are constrained and, at
the tree level, they 
can be written in terms of just two MSSM parameters.
These are commonly chosen to be the mass
of the $CP$-odd neutral Higgs boson, $m_{A^o}$,
and the ratio of the vacuum expectation values
of the two Higgs doublets, $\tan\beta = v_2/v_1$.
The Higgs bosons masses and the mixing angle in the neutral sector are
given in terms of these two parameters, at the tree level, by
\bea
m^2_{H^\pm} &=&  m_{A^o}^2 + m_W^2 \nonumber\\
m^2_{H^{o}, h^{o}} &=& \frac{1}{2}
    \left[ m_{A^o}^2 + m_Z^2 \pm \sqrt{\left(m_{A^o}^2 + m_Z^2\right)^2 -
           4 m_{A^o}^2 m_Z^2 \cos^2 2 \beta }\, \right] \nn \\
 \tan 2 \alpha &=& \tan 2 \beta\, {{m_{A^o}^2 + m^2_Z}\over{m_{A^o}^2 -m^2_Z}}\,.
\eea
The case of large pseudoscalar mass, $m_{A^o} \gg m_{EW}$, deserves
special mention, because in this so-called decoupling
limit~\cite{decoupling} the $h^o$ boson resembles the SM Higgs
boson. 

Once the radiative corrections to the Higgs sector are
included~\cite{loopmassMSSM}, the Higgs boson mass pattern changes 
slightly and depends on the
other MSSM parameters as well. However, irrespectively of the
particular values of the MSSM parameters, there is the important
prediction of the lightest Higgs boson mass being below about $130$
GeV, clearly within the reach of the present and future colliders.

The SUSY-QCD sector in the MSSM consists of squarks and gluinos. We
focus here on the third-generation squarks, which will be the relevant
ones for the dominant radiative corrections to the Yukawa couplings,
and neglect possible intergenerational mixing. 
The tree-level stop and sbottom squared-mass matrices are

\begin{equation}
{\cal M}_{\tilde{t}}^2 =\left(\begin{array}{cc}
M_L^2 &
m_t\, X_t\\ m_t\, X_t & M_R^2
\end{array} \right)\, \, , \, \,
{\cal M}_{\tilde{b}}^2 =\left(\begin{array}{cc}
M_L^{'2} &
m_b\, X_b\\ m_b\, X_b &M_R^{'2} 
\end{array} \right)\,
\label{eq:stopmatrix}
\end{equation}
where
\bea 
M_L^2&=&M_{\tilde Q}^2+m_t^2+\cos{2\beta}(1/2- 2/3 s_W^2)\,m_Z^2 \nonumber \\
M_R^2&=&M_{\tilde U}^2+m_t^2+2/3 \cos{2\beta}\,s_W^2\,m_Z^2 \nonumber \\
X_t&=&A_t-\mu\cot\beta\, \nonumber \\
M_L^{'2}&=&M_{\tilde Q}^2+m_b^2-\cos{2\beta}(1/2- 1/3 s_W^2)\,m_Z^2 \nonumber \\
M_R^{'2}&=&M_{\tilde D}^2+m_b^2\,-1/3 \cos{2\beta}\,s_W^2\,m_Z^2 \nonumber \\
X_b&=&A_b-\mu\tan\beta\,.
\label{eq.MLRtb}
\eea
Here $m_t$ and $m_b$ are the top and bottom quark masses, respectively,
$m_Z$ and $m_W$ are the $Z$ and $W$ gauge boson masses, respectively, 
and $s_W\equiv \sin\theta_W$. The parameters
$M_{\tilde Q}$, $M_{\tilde D}$ and $M_{\tilde U}$ are the soft-SUSY-breaking 
masses
for the third-generation SU(2) squark doublet $(\tilde t_L, \tilde b_L)$
and the singlets $\tilde b_R$ and $\tilde t_R$, respectively.
$\tilde q_{L,R}$ are the superpartners of the chiral projections of
quarks $q_{L,R}=P_{L,R} q$, respectively, where $P_{L,R} = (1 \mp \gamma_5)/2$.
$A_{b,t}$ are the corresponding soft-SUSY-breaking trilinear
couplings and $\mu$ is the bilinear coupling of the two Higgs doublets. 
The squarks mass eigenstates are given by
\begin{equation}
\left( \begin{array}{c}
\tilde q_1\\ \tilde q_2 
\end{array}\right) = (R_q)^{-1} \left( \begin{array}{c}
\tilde q_L\\ \tilde q_R 
\end{array}\right),
\label{eq:squarkrotation}
\end{equation}
where
\begin{equation}
R_q = \left( \begin{array}{cc}
c_q & - s_q \\  s_q & c_q
\end{array}\right), 
\label{eq.rotationmatrix}
\end{equation}
and $c_q=\cos \theta_{q}$, $s_q= \sin \theta_{q}$
with $q=t,b$.
The stop and sbottom mass eigenvalues are given by
\bea
    M^2_{\tilde t_{1,2}} = \frac{1}{2}\left[ M_L^2 + M_R^2
    \pm \sqrt{ (M_L^2 - M_R^2)^2 + 4 m_t^2 X_t^2 } \right]\,,\nonumber\\
 M^2_{\tilde b_{1,2}} = \frac{1}{2}\left[ M_L^{'2} + M_R^{'2}
    \pm \sqrt{ (M_L^{'2} - M_R^{'2})^2 + 4 m_b^2 X_b^2 } \right]\,.
\eea
And the squark mixing angles $\theta_{q}$ $(q=t,b)$ are given by
\bea \label{thetab}
    \cos 2 \theta_{t} = \frac{M_L^2 - M_R^2}
    {M^2_{\tilde t_1}-M^2_{\tilde t_2}}\,&,&\,\,\,\,\,\,
    \cos 2 \theta_{b} = \frac{M_L^{'2} - M_R^{'2}}
    {M^2_{\tilde b_1}-M^2_{\tilde b_2}}, \nonumber \\
    \sin 2 \theta_{q} &=&
    \frac{2 m_q X_q}{M^2_{\tilde q_1}-M^2_{\tilde q_2}}\,.
\eea
 
The free Lagrangian for Higgs bosons, quarks, squarks and Majorana gluinos in
the physical basis are given, respectively, by

\begin{eqnarray}
\mathcal{L}_0 (H) &=& \frac{1}{2} \sum_{H} ( \partial_{\mu} H
\partial^{\mu} H - H m_{H}^2 H) \nn \\
\mathcal{L}_0 (q) &=& \bar b (i\partial{\hspace{-6pt}\slash} - m_b ) b
+ \bar t (i\partial{\hspace{-6pt}\slash} - m_t ) t \nn  \\
\mathcal{L}_0 (\tilde q) &=& \sum_{i=1,2} (\partial_{\mu} \tilde b_i^*
\partial^{\mu}\tilde b_i - M_{\tilde b_i}^2 \tilde b_i^* \tilde b_i) +
\sum_{i=1,2} (\partial_{\mu} \tilde t_i^*
\partial^{\mu}\tilde t_i - M_{\tilde t_i}^2 \tilde t_i^* \tilde t_i)
\nn \\
\mathcal{L}_0 (\tilde g) &=& \frac{1}{2} \bar {\tilde g} (i 
\partial{\hspace{-6pt}\slash}- M_{\tilde g}) \tilde g \, ,
\label{eq.freelagran}
\end{eqnarray}
where the sum in $H$ runs along the Higgs fields
$H=h^o,H^o,A^o,H^1,H^2$, and
$H^1,H^2$ are related to the physical
charged Higgs bosons by $H^1 = (H^+ + H^-)/\sqrt{2}$, $H^2 = -i(H^+ -
H^-)/\sqrt{2}$ and $m_{H^1}^2=m_{H^2}^2=m_{H^\pm}^2$. $M_{\tilde
g}$ is the gluino mass, and we have omitted for brevity
the color indices in the gluinos, quarks and squarks fields.
Notice that in order to make more explicit our posterior assumption
on the mass spectrum, we have denoted by little $m$ all
the light masses of the order of $m_{EW}$, and by capital $M$ all the
heavy masses of the order of $M_{SUSY}$.

The tree-level Yukawa Higgs-quark-quark interactions in the MSSM are
like in a general 2HDMII:
\begin{eqnarray}
\mathcal{L} (H,q) &=& \frac{g m_b \sin \alpha}{2 m_W \cos \beta} h^o \bar b b
- \frac{g m_t \cos \alpha}{2 m_W \sin \beta} h^o \bar t t \nn \\
&-& \frac{g m_b \cos \alpha}{2 m_W \cos \beta} H^o \bar b b
- \frac{g m_t \sin \alpha}{2 m_W \sin \beta} H^o \bar t t \nn \\
&+& \frac{i g m_b}{2 m_W} \tan \beta A^o \bar b \gamma_5 b
+ \frac{i g m_t}{2 m_W} \cot \beta A^o \bar t \gamma_5 t \nn \\
&+& \frac{g m_b}{\sqrt{2} m_W} \tan \beta (H^+ \bar t_L b_R + h.c.)
+ \frac{g m_t}{\sqrt{2} m_W} \cot \beta (H^+ \bar t_R b_L + h.c.) \nn \\
\label{eq.treeyukawa}
\end{eqnarray}

The MSSM gluino-squark-quark interactions terms, in the chiral basis,  are
\begin{equation}
\mathcal{L} (\tilde g, \tilde q, q) = - \sqrt{2} g_s \mathtt{t}
\left( \bar{\tilde g} \tilde t_L^* t_L + \bar{\tilde g} \tilde b_L^* b_L
-  \bar{\tilde g} \tilde t_R^* t_R - \bar{\tilde g} \tilde b_R^* b_R
\right) + h.c 
\label{eq.gluinosqq}
\end{equation}
where $\mathtt{t} \bar {\tilde g} \tilde q^* q \equiv t_{ij}^a \bar
{\tilde g}_a \tilde q^{*i} q^j$, and 
$t_{ij}^a$ are the standard $SU(3)_c$ generators. We will omit the color
indices from now on.

Finally, the Lagrangian terms involving Higgs-squark-squark
interactions, in the chiral squarks basis, are
\begin{eqnarray}
\mathcal{L} (H,\tilde q) &=& \sum_{i,j,H_n} 
\left( (a_{H_n}^{\tilde t})_{ij} 
H_n \tilde t^*_i \tilde t_j + (a_{H_n}^{\tilde b})_{ij} H_n \tilde b^*_i
\tilde b_j \right)
+ \sum_{i,j,H_c} \left( (a_{H_c})_{ij} H_c \tilde t^*_i \tilde b_j
+ h.c. \right) \nn \\
\label{eq.Higgssqsq}
\end{eqnarray}
where the $i$ and $j$ indices run along $L,R$; the Higgs fields $H_n$ along 
$H_n=h^o,H^o,A^o$ and  $H_c$ along 
$H_c=H^1,H^2$. The expressions for the couplings $a_H$ are
given in Appendix A.

\section{Integration of squarks and gluinos}

In this section we perform the integration to one-loop level and to 
$\mathcal{O}(\alpha_s)$ of the squarks and gluinos in the MSSM by
using the standard functional techniques. Since we are studying here
the effect of the SUSY-QCD sector on the Yukawa Higgs-quark-quark
interactions, we take just the relevant terms in the MSSM action,
which can be generically written as
\begin{equation}
S[H, q, \tilde q, \tilde g]=S_0[H] + S_0[q] + S_0[\tilde q]
+S_0[\tilde g]+S[H,q]
+S[\tilde g, \tilde q, q]+S[H,\tilde q],
\end{equation}
where the $S_0$ and $S$ are given in terms of the free and
interaction Lagrangians, respectively, of
Eqs.(\ref{eq.freelagran},\ref{eq.treeyukawa},\ref{eq.gluinosqq}) and
(\ref{eq.Higgssqsq}) by
\begin{equation}
S_0[\Phi]=\int dx {\cal L}_0(\Phi)\, ; \, \Phi=H,q,\tilde q,\tilde g  
\,;\,S[H,q]=\int dx {\cal L}(H,q);
\end{equation}
\begin{equation}
S[\tilde g, \tilde q, q]=\int dx {\cal L}(\tilde g, \tilde q, q)\, ; \,
S[H,\tilde q]=\int dx {\cal L}(H,\tilde q),
\end{equation} 
and $dx$ is the ordinary space-time integration measure.

In the following we introduce some convenient compact notation for the
functional integration.  The 
chiral quark fields, their squarks partners and the physical squarks 
are grouped here as
\begin{eqnarray*}
&Q= 
\left(
\begin{array}{c}
t_L\\ t_R\\b_L\\b_R
\end{array}
\right) \, \, \, , \, \, \,
\tilde Q_{ch} = 
\left(
\begin{array}{c}
\tilde t_L \\ \tilde t_R \\  \tilde b_L \\  \tilde b_R
\end{array}
\right)\, \, \, , \, \, \,
\tilde Q= 
\left(
\begin{array}{c}
\tilde t_1\\ \tilde t_2 \\  \tilde b_1 \\ \tilde b_2
\end{array}
\right).&
\label{eq.quarkandsquark}
\end{eqnarray*}\\
The two squark basis are related via an orthogonal $4 \times 4$ matrix $R$,
given in terms of the matrices $R_q$ ($q=t,b$) of Eq.(\ref{eq.rotationmatrix}), by
\begin{equation}
\tilde Q_{ch} = R \tilde Q \, \, \, , \, \, \, 
R = \left(
\begin{array}{cc}
R_t&0\\ 0&R_b
\end{array}\right). 
\end{equation}

With regard to the gluinos, which are Majorana spinors, we will use a
Majorana representation
for the $\gamma$ Dirac matrices so that the Majorana fields will be
real, $\gamma_i$ ($i=1,2,3$) imaginary and symmetric, $\gamma_0$ and
$\gamma_5$ imaginary and antisymmetric and the chiral projectors
Hermitian.


With this notation, the squark and gluino dependent terms of the above action can be
written in terms of the physical states as follows. The free squark action is
\begin{equation}
S_0[\tilde q]=\int d x \tilde Q^\dagger (-\square-
M_{\tilde q}^2)\tilde Q\equiv <\tilde Q^\dagger A_{\tilde q}^0\tilde Q>,
\end{equation}
where $M_{\tilde q}^2 \equiv diag (M_{\tilde t_1}^2, M_{\tilde t_2}^2,
M_{\tilde b_1}^2,M_{\tilde b_2}^2)$ and $\square \equiv
\partial_{\mu} \partial^{\mu}$. The free action for
the Majorana gluino fields is
\begin{equation}
S_0[\tilde g]=\int d x \frac{1}{2} \bar{\tilde g} (i \partial{\hspace{-6pt}\slash} 
- M_{\tilde g}) \tilde g \equiv \frac{1}{2} <\bar{ \tilde g }
A_{\tilde g}^0 \tilde g>.
\end{equation}
The gluino-squark-quark interactions are written in compact notation as
\begin{equation}
S[\tilde g, \tilde q, q]=\int dx (-\sqrt 2 g_s \mathtt{t} \bar{\tilde
g}\tilde Q^{\dagger}R^T\Sigma Q+h.c.)\equiv <\bar {\tilde g}
\tilde Q^\dagger O Q>+h.c. \, \, ,
\end{equation}
where we have introduced the matrix $\Sigma \equiv diag(1,-1,1,-1)$ and
the interaction operator is defined as
\begin{equation}
O \equiv -\sqrt 2 g_s \mathtt{t} R^T \Sigma.
\label{odefinition}
\end{equation}
The Higgs-squark-squark interactions are given by\footnote{Since we are not
going to integrate here the Higgs fields, we work in the unitary gauge
and refer always to the physical Higgs basis.}
\begin{equation}
S[H,\tilde q]=\sum_{H} \int dx H \tilde Q^\dagger B_H \tilde
Q\equiv <H \tilde Q^\dagger B_H \tilde Q>,
\end{equation}
where $H$ runs again along $H=h^o,H^o,A^o,H^1,H^2$, $B_H \equiv R^T A_H R$ and $A_H$
 are the $4 \times 4$ coupling matrices that are defined in Appendix A.

We next outline the computation of the contribution to the quarks and
Higgs bosons effective action coming from the functional integration of the squark and
gluino fields, $\Delta \Gamma_{eff}[H,q]$. This integration can be written as
\begin{equation}
e^{i\Delta \Gamma_{eff}[H,q]}=\int [d\tilde Q^\dagger][d\tilde Q ]
[d\tilde g]e^{i(S_0[\tilde q]
+S_0[\tilde g]
+S[\tilde g, \tilde q, q]+S[H,\tilde q])}.
\label{eq.defgammaeff}
\end{equation}
After this integration the total effective action for quarks and Higgs
particles will then be obtained by
\begin{equation}
\Gamma_{eff} [H,q] = S_0[H]+ S_0[q] +  S[H,q] + \Delta \Gamma_{eff}[H,q].
\label{eq.effaction}
\end{equation}
In order to perform first the gluino integration we write
Eq.(\ref{eq.defgammaeff}) as
\begin{equation}
e^{i\Delta \Gamma_{eff}[H,q]}=\int [d\tilde Q^\dagger][d\tilde Q]
e^{i<\tilde Q^\dagger (A_{\tilde q}^0+H B_H)\tilde Q>} \int
[d\tilde g]e^{i\frac{1}{2}  <\bar{ \tilde g } A_{\tilde g}^0 g> + i(<\bar {\tilde g}
\tilde Q^\dagger O Q>+h.c.)}.
\end{equation}
A standard Gaussian integration of the real  Grassmann 
variable $\tilde g$ yields
\begin{equation}
e^{i\Delta \Gamma_{eff}[H,q]}=\int [d\tilde Q^\dagger][d\tilde Q]
e^{i<\tilde Q^\dagger (A_{\tilde q}^0+H B_H)\tilde
Q>}e^{-\frac{i}{2}<\bar \eta(A_{\tilde g}^0)^{-1}\eta>},
\end{equation}
where we have omitted a field independent global factor that is irrelevant for the
effective action, and we have introduced the Majorana field
\begin{equation}
\eta \equiv \tilde Q^\dagger O Q + Q^\dagger O^\dagger \tilde Q.
\end{equation}
Thus, we have
\begin{eqnarray}
e^{i\Delta \Gamma_{eff}[H,q]} = \int [d\tilde Q^\dagger][d\tilde Q] 
e^{i<\tilde Q^\dagger (A_{\tilde q}^0+H B_H-O  Q
\gamma_0 D_{\tilde g}Q^\dagger O^\dagger )\tilde Q>}&& \nonumber \\
= [det(A_{\tilde q}^0+H B_H-O  Q \gamma_0 D_{\tilde
g}Q^\dagger O^\dagger )]^{-1},&&
\end{eqnarray}
where we have performed another complex-scalar Gaussian integration
and $D_{\tilde g}=(A_{\tilde g}^0)^{-1}$ is the gluino propagator
which, in the position representation, can be written as
\begin{equation}
D_{\tilde g xy}=\int d \tilde k \frac{e^{-ik(x-y)}}{k{\hspace{-6pt}\slash}-
M_{\tilde g}},
\label{eq.propgluino}
\end{equation}
where $d \tilde k \equiv d^4k/(2\pi)^4$. Therefore, aside from a field
independent term, the contribution to the effective
action  can be written as
\begin{equation}
\Delta \Gamma_{eff}[H,q]=i Tr \log[1 + D_{\tilde q}(H B_H-O Q
\gamma_0 D_{\tilde g}Q^\dagger O^\dagger )],
\label{eq.efflog}
\end{equation}
where $Tr$ is the functional trace, and the squark propagator
$D_{\tilde q}= (A_{\tilde q}^0)^{-1}$ is
given, in the position representation, by
\begin{equation}
D_{\tilde q xy}=\int d \tilde q \frac{e^{-iq(x-y)}}{q^2-
M_{\tilde q}^2},
\label{eq.propsquark}
\end{equation}
where $\frac{1}{q^2-M_{\tilde q}^2} \equiv diag(\frac{1}{q^2-M_{\tilde t_1}^2},
\frac{1}{q^2-M_{\tilde t_2}^2}, \frac{1}{q^2-M_{\tilde b_1}^2},
\frac{1}{q^2-M_{\tilde b_2}^2})$.

Expanding the logarithm in Eq.(\ref{eq.efflog}) we get
\begin{equation}
\Delta \Gamma_{eff}[H,q]=i \sum_{k=1}^\infty \frac{(-)^{k+1}}{k}
Tr [ D_{\tilde g}(H B_H-O Q \gamma_0 D_{\tilde g}Q^\dagger
O^\dagger )]^k\equiv  \sum_{k=1}^\infty \Delta
\Gamma_{eff}^{(k)}
\label{eq.neff}
\end{equation}

In this work we are interested in the possible one-loop
effects of the SUSY-QCD sector on the Higgs couplings to quark pairs
and, therefore, only a few terms of this expansion will be taken
into account. For $k=1$ we have
\begin{equation}
\Delta \Gamma_{eff}^{(1)}=i Tr D_{\tilde q}  (H B_H-O Q \gamma_0
D_{\tilde g}Q^\dagger O^\dagger),
\end{equation}
which in the position representation reads
\begin{equation}
\Delta \Gamma_{eff}^{(1)}=i tr \sum_H \int dx  D_{\tilde q xx} (H_x B_H)
-i tr \int dx dy D_{\tilde q xy}O Q_y \gamma_0 D_{\tilde g
yx}Q^\dagger_x O^\dagger ,
\end{equation}
where the notation for the fields is $H_x = H(x)$, $Q_y=Q(y)$, etc.
The first term corresponds to the Higgs tadpoles with squark loops
which are of zero order in
$\alpha_s$ and, hence, will not be considered in the following. The
second term gives the one-loop contributions to the quark propagators
which are of order $\alpha_s$ and involves both a gluino and a squark
propagator in the loop. By substituting the previous expressions for
the operator $O$ and for the squark and gluino propagators, and after
some algebra, this second term can be written as
\begin{equation}
\Delta \Gamma_{eff}^{(1)}=\frac{2 \alpha_s}{3\pi} \int dx dy \sum_{abc}
 \bar Q_x^c (\Sigma R)_{ca} \Pi_{xy}^a (R^T \Sigma)_{ab} Q_y^b,
\label{eq.effres1}
\end{equation}
where
\begin{equation}
\Pi_{xy}^a \equiv \int d\tilde p e^{-ip(x-y)} [p{\hspace{-6pt}\slash}
B_1 + M_{\tilde g} B_0] (p^2, M_{\tilde g}^2, M_{\tilde q_a}^2)
\label{eq.twopoint}
\end{equation}
and the indices $a,b,c$ run along the four entries corresponding to the
quarks and the (physical) squarks matrices in Eq.(\ref{eq.quarkandsquark}).

The $k=2$ contribution is given by
\begin{equation}
\Delta \Gamma_{eff}^{(2)}=-\frac{i}{2}Tr[ D_{\tilde q}  (H B_H-O Q
\gamma_0 D_{\tilde g}Q^\dagger O^\dagger)]^2=i Tr (H B_H D_{\tilde
q} O Q \gamma_0 D_{\tilde g}Q^\dagger O^\dagger D_{\tilde q}) 
\end{equation}
where in the last equality we have omitted those terms that do not
contribute to the Higgs-quark-quark 
couplings. In the position representation this last term can be written as
\begin{equation}
\Delta \Gamma_{eff}^{(2)}=i tr \sum_H \int dx dy dz H_x B_H D_{\tilde q
xy} O Q_y\gamma_0 D_{\tilde g y z}Q^\dagger_z O^\dagger
D_{\tilde q zx}.
\label{eq.eff2}
\end{equation}
Finally, by substituting the operator $O$ and the squark and gluino
propagators of Eqs.(\ref{odefinition},\ref{eq.propgluino}) and
(\ref{eq.propsquark}) in Eq.(\ref{eq.eff2}),
and after some algebra, this contribution to the effective action can be written in
terms of the standard one-loop integrals as
\begin{equation}
\Delta \Gamma_{eff}^{(2)}=-\frac{2 \alpha_s}{3\pi} \sum_H \int dx dy dz
\sum_{abcd} \bar Q_z^d (\Sigma R)_{da} H_x B_H^{ab} V_{xyz}^{ab} (R^T
\Sigma)_{bc} Q_y^c,
\label{eq.effres2}
\end{equation}
where
\begin{equation}
 V_{xyz}^{ab} \equiv \int d\tilde q d\tilde p e^{-iq(x-y)}
 e^{-ip(z-y)} [p{\hspace{-6pt}\slash} C_{11} + q{\hspace{-6pt}\slash}
 C_{12} + M_{\tilde g} C_0 ] 
(p^2,q^2,r^2,M_{\tilde g}^2,M_{\tilde q_a}^2,M_{\tilde q_b}^2)
\label{eq.threepoint}
\end{equation}
with $r \equiv -(p+q)$.
Here again, the indices $a,b,c,d$ run along the four entries of the
quark and squark matrices and $H$ runs along the Higgs fields
$H=h^o,H^o,A^o,H^1,H^2$. The last parenthesis refers to the arguments
of the scalar three-point integrals. Our definition of the scalar one-loop
integrals $B_0,B_1,C_0,C_{11}$ and $C_{12}$ is as in~\cite{Hollik} and is
given for completeness in Appendix B.

\begin{figure}
\begin{center}
\epsfig{file=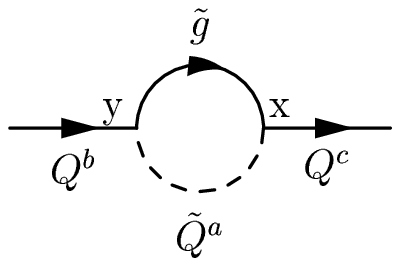,width=5cm}~~
\epsfig{file=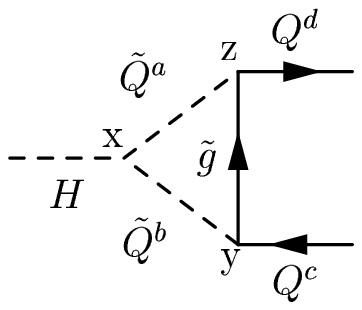,width=5cm}\vspace*{1.2cm}
\caption{\it Diagrams corresponding to $\Delta \Gamma_{eff}^{(1)}$ and 
$\Delta \Gamma_{eff}^{(2)}$ of Eqs.(\ref{eq.effres1},\ref{eq.twopoint}) and
(\ref{eq.effres2},\ref{eq.threepoint}), respectively.}
\label{fig:diagramas}
\end{center}
\end{figure}

The corresponding generic Feynman diagrams for $\Delta \Gamma_{eff}^{(1)}$ and 
$\Delta \Gamma_{eff}^{(2)}$ of Eqs.(\ref{eq.effres1},\ref{eq.twopoint}) and
(\ref{eq.effres2},\ref{eq.threepoint}), respectively, are shown in Fig.1.

\section{Effective Higgs-quark-quark interactions from
heavy squarks and gluinos}

In this section we compute the effective local interactions of Higgs
particles with quarks to one-loop level and $\mathcal{O}(\alpha_s)$, in
the limit of heavy squarks and gluinos. We derive
the quarks and Higgs bosons effective Lagrangian starting 
from the nonlocal effective action of
Eqs.(\ref{eq.effaction},\ref{eq.neff},\ref{eq.effres1},\ref{eq.twopoint},\ref{eq.effres2})
and (\ref{eq.threepoint})
which includes the contributions coming from integrating out the squarks and
gluino fields. Next we perform a large $M_{SUSY}$
expansion that is valid in the limit
of nearly degenerate and very heavy SUSY particles, 
$M_{\tilde g} \sim M_{\tilde q} \sim M_{SUSY} \gg m_{EW}$,
where $m_{EW}$ stands for any mass or external momentum associated with a
non-SUSY particle, i.e. quarks, gauge bosons and Higgs particles. 
For that purpose, we consider all
the SUSY mass parameters and the $\mu$ parameter of the same order (collectively denoted by
$M_{SUSY}$) and much heavier than the electroweak
scale\footnote{Notice that our choice of large $\mu$ is not necessary
for a heavy SUSY-QCD spectrum but it will be mandatory to get all the
gauginos heavy in the SUSY-electroweak sector. Although it is not
needed for the present work, we are thinking of the whole SUSY
spectrum being heavy as compared to the electroweak scale.}, 
that is, $M_{SUSY} \sim M_{\tilde g} \sim  M_{\tilde Q} \sim 
M_{\tilde U} \sim  M_{\tilde D} \sim \mu  \gg m_{EW}$.

The effective Lagrangian is then defined in the usual way,
\begin{equation}
\Gamma_{eff} [H,q] \equiv \int dx \mathcal{L}_{eff} (H,q),
\end{equation}
and, correspondingly to Eq.(\ref{eq.effaction}), it can be written as follows
\begin{equation}
\mathcal{L}_{eff} (H,q) =\mathcal{L}_0 (H) + \mathcal{L}_{0}
(q) + \mathcal{L} (H,q) + \Delta \mathcal{L}_{eff} (H,q). 
\end{equation}

In order to compute $\Delta \mathcal{L}_{eff} (H,q)$ to one-loop order,
$\mathcal{O}(\alpha_s)$, and in the limit of heavy squarks and gluinos,
we substitute into
Eqs.(\ref{eq.effres1},\ref{eq.twopoint},\ref{eq.effres2}) and
(\ref{eq.threepoint}) the corresponding interaction matrices $\Sigma,
R$ and $B_H$ and the expressions for the large mass expansion of the one-loop
integrals and mixing angles given in Appendix B. 
After some algebra, we obtain the following result
\begin{eqnarray}
\Delta \mathcal{L}_{eff} (H,q) &=& \frac{\alpha_s}{3 \pi} \Delta \,
\, \bar b
(i \partial{\hspace{-6pt}\slash}) b 
+ \frac{\alpha_s}{3 \pi} \frac{M_{\tilde g} (A_b - \mu
\tan\beta)}{M_{SUSY}^2} \, m_b \bar b b
\nn \\
&&+ \frac{\alpha_s}{3 \pi} \Delta \,
\, \bar t
(i \partial{\hspace{-6pt}\slash}) t 
+ \frac{\alpha_s}{3 \pi} \frac{M_{\tilde g} (A_t - \mu \cot
\beta)}{M_{SUSY}^2} \, m_t \bar t t \nn \\
&&- \frac{\alpha_s}{3 \pi} \frac{g m_b \sin\alpha}{2 m_W \cos\beta}
\frac{M_{\tilde g} (A_b + \mu \cot\alpha)}{M_{SUSY}^2} h^o \bar b b
\nn \\
&&+ \frac{\alpha_s}{3 \pi} \frac{g m_t \cos\alpha}{2 m_W \sin\beta}
\frac{M_{\tilde g} (A_t + \mu \tan\alpha)}{M_{SUSY}^2} h^o \bar t t
\nn \\
&&+ \frac{\alpha_s}{3 \pi} \frac{g m_b \cos\alpha}{2 m_W \cos\beta}
\frac{M_{\tilde g} (A_b - \mu \tan\alpha)}{M_{SUSY}^2} H^o \bar b b
\nn \\
&&+ \frac{\alpha_s}{3 \pi} \frac{g m_t \sin\alpha}{2 m_W \sin\beta}
\frac{M_{\tilde g} (A_t - \mu \cot\alpha)}{M_{SUSY}^2} H^o \bar t t
\nn \\
&&- \frac{\alpha_s}{3 \pi} \frac{i g m_b}{2 m_W} \tan\beta
\frac{M_{\tilde g} (A_b + \mu \cot\beta)}{M_{SUSY}^2} A^o \bar b \gamma_5 b
\nn \\
&&- \frac{\alpha_s}{3 \pi} \frac{ig m_t}{2 m_W} \cot\beta
\frac{M_{\tilde g} (A_t + \mu \tan\beta)}{M_{SUSY}^2} A^o \bar t
\gamma_5 t
\nn \\
&&- \frac{\alpha_s}{3 \pi} \frac{g m_b}{\sqrt{2} m_W} \tan\beta
\frac{M_{\tilde g} (A_b + \mu \cot\beta)}{M_{SUSY}^2} (H^+ \bar t_L
b_R + h.c.) \nn \\
&&- \frac{\alpha_s}{3 \pi} \frac{g m_t}{\sqrt{2} m_W} \cot\beta
\frac{M_{\tilde g} (A_t + \mu \tan\beta)}{M_{SUSY}^2} (H^+ \bar t_R
b_L + h.c.) \nn \\
&&+ \mathcal{O} \left(\frac{m_{EW}}{M_{SUSY}} \right)^n
\label{eq.extralag}
\end{eqnarray}

Some comments are in order. First notice that, as expected, the
 SUSY-QCD effects summarized in Eq.(\ref{eq.extralag}) include both the divergent
and the finite contributions from heavy squarks and gluinos.
We have used dimensional regularization to
perform the one-loop integrals with $\Delta \equiv 2/\epsilon - \gamma_E +
\log(4\pi) - log (M_{SUSY}^2/\mu_0^2)$, $\epsilon \equiv 4 -D$ and
 $\mu_0$ is the usual mass scale of dimensional
 regularization\footnote{Notice that, presumably, the same
results would have been obtained by using instead dimensional
 reduction.}. 
Second, all these leading terms are of
 $\mathcal{O}(m_{EW}/M_{SUSY})^0$, apart from the logarithms, 
in the large $M_{SUSY}$ expansion.
The last contribution in Eq.(\ref{eq.extralag}),  not shown
explicitly, refers to the next to leading order terms in the large
$M_{SUSY}$ expansion which are suppressed by inverse powers of
$M_{SUSY}$ ($n>0$) and vanish in the asymptotic $M_{SUSY} \to
\infty$ limit. It is worth mentioning here that, with this effective
action formalism, one can easily get as well these next to leading
order contributions by simply including the next terms in the
expansions of the one-loop integrals and mixing angles. We will not
work out here these terms but concentrate just on the leading terms
 which are the only ones that remain in the asymptotic large
 $M_{SUSY}$ limit. For a
discussion on the next to leading order terms we refer the reader 
to~\cite{HaberTemes,ourHtb} where they have been obtained for the 
$h^o \bar bb$ and $H^+ \bar t b$ couplings, using a Feynman
 diagrammatic approach.

In order to study the decoupling behavior of the SUSY-QCD sector in
the Yukawa Higgs-quark-quark couplings we have to find out if the
effects found in Eq.(\ref{eq.extralag}) can be absorbed or not into
redefinitions of the tree-level masses, couplings and wave functions
of the light fields, namely, $H$ and $q$. We first redefine the quark
fields and masses such that their corresponding kinetic and mass terms
get the canonical form
\begin{eqnarray}
b &=& \left( 1 - \frac{\alpha_s}{6 \pi} \, \, \Delta \right) b' \nn \\      
t &=& \left( 1 - \frac{\alpha_s}{6 \pi} \, \, \Delta \right) t' \nn \\
m_b &=& \left ( 1 + \frac{\alpha_s}{3 \pi} \, \, \Delta +
\frac{\alpha_s}{3 \pi} \frac{M_{\tilde g} (A_b - \mu \tan\beta)}{M_{SUSY}^2}
\right) m'_b \nn \\
m_t &=& \left ( 1 + \frac{\alpha_s}{3 \pi} \, \, \Delta +
\frac{\alpha_s}{3 \pi} \frac{M_{\tilde g} (A_t - \mu \cot\beta)}{M_{SUSY}^2}
\right) m'_t.
\end{eqnarray}

After this, the remaining contributions in $\Delta \mathcal{L}_{eff}$
are rewritten in terms of these new fields $q'$ and masses $m'_q$, and we finally
get
\begin{eqnarray}
\mathcal{L}_{eff} (H,q') &=& \mathcal{L}_0 (H) 
+ \bar b' (i\partial{\hspace{-6pt}\slash} - m'_b ) b'
+ \bar t' (i\partial{\hspace{-6pt}\slash} - m'_t ) t' \nn \\ 
&&+ \frac{g m'_b \sin \alpha}{2 m_W \cos \beta} \left[1 
-\frac{\alpha_s}{3 \pi} \frac{M_{\tilde g} \mu}{M_{SUSY}^2}
(\tan\beta+ \cot\alpha) \right] h^o \bar b' b' \nn \\
&&- \frac{g m'_t \cos \alpha}{2 m_W \sin \beta} \left[1 -
\frac{\alpha_s}{3 \pi} \frac{M_{\tilde g} \mu}{M_{SUSY}^2} (\cot\beta
+ \tan\alpha) \right] h^o \bar t' t' \nn \\
&&- \frac{g m'_b \cos \alpha}{2 m_W \cos\beta} \left[1 -
\frac{\alpha_s}{3 \pi} \frac{M_{\tilde g} \mu}{M_{SUSY}^2} (\tan\beta
- \tan\alpha) \right] H^o \bar b' b' \nn \\
&&- \frac{g m'_t \sin \alpha}{2 m_W \sin\beta} \left[1 -
\frac{\alpha_s}{3 \pi} \frac{M_{\tilde g} \mu}{M_{SUSY}^2} (\cot\beta
- \cot\alpha) \right] H^o \bar t' t' \nn \\
&&+ \frac{i g m'_b}{2 m_W} \tan\beta  \left[1 -
\frac{\alpha_s}{3 \pi} \frac{M_{\tilde g} \mu}{M_{SUSY}^2} (\tan\beta
+ \cot\beta) \right] A^o \bar b' \gamma_5 b' \nn \\
&&+ \frac{i g m'_t}{2 m_W} \cot\beta  \left[1 -
\frac{\alpha_s}{3 \pi} \frac{M_{\tilde g} \mu}{M_{SUSY}^2} (\tan\beta
+ \cot\beta) \right] A^o \bar t' \gamma_5 t' \nn \\
&&+ \frac{g m'_b}{\sqrt{2} m_W} \tan\beta  \left[1 -
\frac{\alpha_s}{3 \pi} \frac{M_{\tilde g} \mu}{M_{SUSY}^2} (\tan\beta
+ \cot\beta) \right] (H^+ \bar t'_L b'_R + h.c.) \nn \\
&&+ \frac{g m'_t}{\sqrt{2} m_W} \cot\beta  \left[1 -
\frac{\alpha_s}{3 \pi} \frac{M_{\tilde g} \mu}{M_{SUSY}^2} (\tan\beta
+ \cot\beta) \right] (H^+ \bar t'_R b'_L + h.c.) \nn \\
\end{eqnarray}
Notice that, as expected, the Higgs fields and Higgs masses do
not get corrections from the SUSY-QCD sector at the one-loop level and 
$\mathcal{O}(\alpha_s)$, while the quark fields and quark masses do get
corrections from this sector. Notice also that the final result is
independent of the trilinear soft-breaking parameters, $A_{t,b}$.

The most relevant result is that we have
found new Yukawa interactions which were not present in the tree-level
interaction Lagrangian, $\mathcal{L} (H,q)$, and that cannot be
absorbed into redefinitions of the masses, couplings and wave
functions of the light fields.  In order to
illustrate this more clearly, we rewrite $\mathcal{L}_{eff}$ in terms
of a set of new Yukawa couplings, $\Delta \lambda$,
\begin{eqnarray}
\mathcal{L}_{eff} (H,q) &=& \mathcal{L}_0 (H) + \mathcal{L}_0 (q) + \mathcal{L}(H,q) 
+ \Delta \lambda^{h^o\bar bb} h^o \bar b b +  \Delta \lambda^{h^o\bar tt} h^o 
\bar t t \nn \\
&& + \Delta \lambda^{H^o\bar bb} H^o \bar b b + \Delta \lambda^{H^o\bar tt} 
H^o \bar t t 
+ \Delta \lambda^{A^o\bar bb} A^o \bar b b +
\Delta \lambda^{A^o\bar tt} A^o \bar t t \nn \\
&&+ \Delta \lambda^{H^+\bar t_L b_R} (H^+ \bar t_L b_R + h.c. )
+ \Delta \lambda^{H^+ \bar t_R b_L} (H^+ \bar  t_R b_L + h.c. ) \nn \\
\label{eq.newlagran}
\end{eqnarray}

where
\begin{eqnarray}
\Delta \lambda^{h^o\bar bb} &\equiv& \frac{g m_b \sin \alpha}{2 m_W \cos \beta} \left[-
\frac{\alpha_s}{3 \pi} \frac{M_{\tilde g} \mu}{M_{SUSY}^2} (\tan\beta
+ \cot\alpha) \right] \nn \\
\Delta \lambda^{h^o\bar tt} &\equiv& - \frac{g m_t \cos \alpha}{2 m_W \sin \beta} \left[-
\frac{\alpha_s}{3 \pi} \frac{M_{\tilde g} \mu}{M_{SUSY}^2} (\cot\beta
+ \tan\alpha) \right] \nn \\
\Delta \lambda^{H^o\bar bb} &\equiv& - \frac{g m_b \cos \alpha}{2 m_W \cos\beta} \left[-
\frac{\alpha_s}{3 \pi} \frac{M_{\tilde g} \mu}{M_{SUSY}^2} (\tan\beta
- \tan\alpha) \right] \nn \\
\Delta \lambda^{H^o\bar tt} &\equiv& - \frac{g m_t \sin \alpha}{2 m_W \sin\beta} \left[-
\frac{\alpha_s}{3 \pi} \frac{M_{\tilde g} \mu}{M_{SUSY}^2} (\cot\beta
- \cot\alpha) \right] \nn \\
\Delta \lambda^{A^o\bar bb} &\equiv& \frac{i g m_b}{2 m_W} \tan\beta  \left[-
\frac{\alpha_s}{3 \pi} \frac{M_{\tilde g} \mu}{M_{SUSY}^2} (\tan\beta
+ \cot\beta) \right] \nn \\
\Delta \lambda^{A^o\bar tt} &\equiv& \frac{i g m_t}{2 m_W} \cot\beta  \left[-
\frac{\alpha_s}{3 \pi} \frac{M_{\tilde g} \mu}{M_{SUSY}^2} (\tan\beta
+ \cot\beta) \right] \nn \\
\Delta \lambda^{H^+\bar t_L b_R} &\equiv& \frac{g m_b}{\sqrt{2} m_W} \tan\beta  \left[-
\frac{\alpha_s}{3 \pi} \frac{M_{\tilde g} \mu}{M_{SUSY}^2} (\tan\beta
+ \cot\beta) \right] \nn \\
\Delta \lambda^{H^+\bar t_R b_L} &\equiv& \frac{g m_t}{\sqrt{2} m_W} \cot\beta  \left[-
\frac{\alpha_s}{3 \pi} \frac{M_{\tilde g} \mu}{M_{SUSY}^2} (\tan\beta
+ \cot\beta) \right] 
\label{eq.newcoup}
\end{eqnarray}
and we have omitted the primes in the quark fields and masses for brevity.
The previous result in Eqs(\ref{eq.newlagran},\ref{eq.newcoup}) is an explicit
demonstration of non-decoupling of heavy squarks and gluinos in the
low energy Yukawa Higgs-quark-quark interactions.

\section{Discussion and conclusions}

The new Yukawa couplings that have been presented in
Eqs.(\ref{eq.newlagran},\ref{eq.newcoup}) summarize all the
non-decoupling SUSY-QCD effects, to one-loop and
$\mathcal{O}(\alpha_s)$ from nearly degenerate heavy squarks and
gluinos, in the low energy Higgs-quark-quark interactions. This
non-decoupling behavior can be seen in that the new low energy
effective interactions do remain even in the extremely heavy SUSY
masses limit, $M_{\tilde q} \sim M_{\tilde g} \sim \mu \sim M_{SUSY}
\gg m_{EW}$. Furthermore, by comparison of our result with the
Lagrangian for Higgs-quark-quark interactions in a general 2HDM we
realize that the complete set of Yukawa interactions 
are exactly those of a 2HDMIII with the particular
values of the Yukawa couplings given by Eqs.(\ref{eq.treeyukawa},\ref{eq.newlagran})
and (\ref{eq.newcoup}). Thus, as announced, one starts in the MSSM at
tree level with a Higgs sector of type 2HDMII and, after integration 
of squarks and
gluinos, one ends up with an unconstrained 2HDMIII. The
track of SUSY has not been lost but it is present in the low energy
2HDMIII interactions via the specific values of the Yukawa couplings,
which in turn depend on the underlying MSSM parameters.

Notice that all the previous results are valid for any value of the
input parameters in the MSSM Higgs sector, $m_{A^o}$ and
$\tan\beta$. There are two particularly interesting situations that
are worth mentioning: the large $m_{A^o} \gg m_{EW}$ limit, defined as
the decoupling limit in~\cite{decoupling}, and the large
$\tan\beta \gg 1$ limit. In the former limit, $\cot\alpha$
approaches $-\tan\beta$ (and, correspondingly, 
$\tan\alpha \rightarrow -\cot\beta$) and, as can be seen 
in Eq.({\ref{eq.newcoup}}), the SUSY-QCD contributions decouple in
$\Delta \lambda^{h^obb}$ and $\Delta \lambda^{h^ott}$ and, therefore,
the $h^o$ Yukawa couplings to quarks cannot be distinguished from the
Higgs-quark-quark couplings in the SM. 
The other new couplings get exactly the same non-decoupling 
factor, being proportional to $(\tan\beta+\cot\beta)$. 
However, this non-decoupling behavior is not expected to leave
observable signals since, in this limit, the masses of $H^o$, $A^o$
and $H^{\pm}$ also get very heavy, $m_{H^o} \simeq m_{H^{\pm}} \simeq
m_{A^o} \gg m_{EW}$ and they will probably not be reachable in the
forthcoming colliders. The phenomenology of the MSSM Higgs sector will
be just as in the SM, with $h^o$ being undistinguisable from the SM
Higgs boson.

On the other hand, in the large $\tan \beta$ limit (with $m_{A^o}$
fixed) there are some corrections $\Delta\lambda/\lambda_0$ that grow
linearly with $\tan\beta$ and can be numerically large for large
values of $\tan\beta$.
As can be seen in Eq.(\ref{eq.newcoup}), these corrections, which are enhanced
by $\tan\beta$ factors, are just present in $\Delta \lambda^{h^o\bar b
b}$, $\Delta \lambda^{H^o\bar b b}$, $\Delta \lambda^{A^o\bar b b}$,
$\Delta \lambda^{A^o\bar t t}$ and $\Delta \lambda^{H^+\bar t b}$, in
agreement with previous partial results
~\cite{cmwpheno,polonsky,kolda,CarenaDavid,EberlTodos,HaberTemes,ourHtb}.

The results presented in this work are relevant for their phenomenological implications in
Higgs production and decays at present and future
colliders. Particularly interesting are the modifications they imply
on the predictions for the Higgs bosons and top quark widths. By simply
replacing the tree-level Yukawa couplings $\lambda_{0}^{H\bar qq'}$ 
($\lambda_{0}^{h^o \bar b b} = gm_b\sin\alpha/2m_W\cos\beta$, etc.)
by the corresponding effective couplings, $\lambda_{eff}^{H\bar qq'} \equiv
\lambda_{0}^{H\bar qq'} + \Delta \lambda^{H\bar qq'}$, with 
$\Delta \lambda^{H\bar qq'}$
given in Eq.(\ref{eq.newcoup}), we find out the following results for
the one-loop, $\mathcal{O}(\alpha_s)$ SUSY-QCD contributions to the
partial decay widths:
\begin{eqnarray}
\Gamma (h^0 \to \bar b b) =\Gamma_0 (h^0 \to \bar b b)  \left[ 1 -
\frac{2\alpha_s}{3 \pi} \frac{M_{\tilde g} \mu}{M_{SUSY}^2} (\tan\beta
+ \cot\alpha) \right] \nn \\
\Gamma (H^0 \to \bar b b) =\Gamma_0 (H^0 \to \bar b b) \left[ 1 -
\frac{2\alpha_s}{3 \pi} \frac{M_{\tilde g} \mu}{M_{SUSY}^2} (\tan\beta
- \tan\alpha) \right] \nn \\
\Gamma (H^0 \to \bar t t) =\Gamma_0 (H^0 \to \bar t t)\left[1 -
\frac{2\alpha_s}{3 \pi} \frac{M_{\tilde g} \mu}{M_{SUSY}^2} (\cot\beta
- \cot\alpha) \right] \nn \\
\Gamma (A^0 \to \bar b b) =\Gamma_0 (A^0 \to \bar b b)\left[1 -
\frac{2\alpha_s}{3 \pi} \frac{M_{\tilde g} \mu}{M_{SUSY}^2} (\tan\beta
+ \cot\beta) \right] \nn \\
\Gamma (A^0 \to \bar t t) =\Gamma_0 (A^0 \to \bar t t)\left[1 -
\frac{2\alpha_s}{3 \pi} \frac{M_{\tilde g} \mu}{M_{SUSY}^2} (\tan\beta
+ \cot\beta) \right] \nn \\
\Gamma (H^+ \to t \bar b) =\Gamma_0 (H^+ \to t \bar b)\left[1 -
\frac{2\alpha_s}{3 \pi} \frac{M_{\tilde g} \mu}{M_{SUSY}^2} (\tan\beta
+ \cot\beta) \right] \nn \\
\Gamma (t \to H^+ b) =\Gamma_0 (t \to H^+ b) \left[1 -
\frac{2\alpha_s}{3 \pi} \frac{M_{\tilde g} \mu}{M_{SUSY}^2} (\tan\beta
+ \cot\beta) \right] 
\end{eqnarray}
where $\Gamma_0$ are the tree-level widths. These formulas are for
nearly degenerate heavy squarks and gluinos and apply for any value of
$m_{A^o}$ and $\tan\beta$. They are in perfect agreement with the
expressions obtained from a complete diagrammatic computation of the
partial widths (see, for instance,~\cite{HaberTemes,ourHtb} for the $h^o$
and $H^+$ cases).
Notice that, in most of the MSSM parameter
space, these SUSY-QCD corrections have all the same sign and universal
character, specially in the large $\tan\beta$ regime, and they will be very
useful in a global experimental analysis,
because they will yield strongly correlated signals in event rates from
Higgs (and top) decays~\cite{curiel}. These corrections can be used
for indirect
but very efficient SUSY searches at the present and future 
colliders~\cite{cmwpheno,kolda,CarenaDavid,EberlTodos,polonsky,HaberTemes,ourHtb,RADCOR2000,HaberLC,MJsitges,siannah,curiel}.

In summary, in this work we have computed the SUSY-QCD
contributions to all the Higgs-quark-quark couplings at the one-loop
level and $\mathcal{O}(\alpha_s)$ in the limit of very heavy squarks
and gluinos. We have found that, after integrating out
the squark and gluino fields, these contributions do not decouple in
the Higgs-quark-quark interactions, and give as a result 
a low energy effective theory of type 2HDMIII,
 with specific values for the couplings given in terms of the
underlying MSSM parameters.

\section*{Acknowledgments}
This work has been supported in part by
the Spanish Ministerio de Ciencia y Tecnolog{\'\i}a under projects CICYT
FPA 2000-0980, CICYT AEN 97-1693 and PB98-0782.


\section*{Appendices}
\vspace{0.4cm}
\setcounter{equation}{0}

\section*{{\bf{A}} Higgs-squark-squark  interaction Lagrangian}
\renewcommand{\theequation}{A.\arabic{equation}}

The interactions between Higgs particles and squarks, in the chiral
 squark basis,  are described by
the following interaction Lagrangian:

\begin{equation}
\mathcal{L} (H, \tilde q) = \sum_{H} H \tilde Q_{ch}^{\dagger}
A_H \tilde Q_{ch} 
\end{equation}
where the $H$ index runs along $H=h^o,H^o,A^o,H^1,H^2$.
In this appendix we give the expressions for the coupling $4 \times 4$
matrices $A_H$ corresponding to the different Higgs fields.

For $h^o$
\begin{equation}
A_{h^o} = \left( \begin{array}{cc}
a_{h^o}^{\tilde t} & 0 \\ 0 & a_{h^o}^{\tilde b} 
\end{array}\right) 
\end{equation}

where

\begin{eqnarray}
a_{h^o}^{\tilde t} &=& \left( \begin{array}{cc}
\frac{g m_Z}{c_W}(\frac{1}{2} - \frac{2}{3} s_W^2) \sin (\alpha + \beta) 
- \frac{g m_t^2 \cos \alpha}{m_W \sin \beta} &
-\frac{g m_t}{2 m_W \sin \beta} (\mu \sin \alpha + A_t \cos \alpha) \\
-\frac{g m_t}{2 m_W \sin \beta} (\mu \sin \alpha + A_t \cos \alpha) &
\frac{g m_Z}{c_W}\, \frac{2}{3} s_W^2 \sin (\alpha + \beta) 
- \frac{g m_t^2 \cos \alpha}{m_W \sin \beta}
\end{array}\right) \nn \\
a_{h^o}^{\tilde b} &=& \left( \begin{array}{cc}
\frac{g m_Z}{c_W}(- \frac{1}{2} + \frac{1}{3} s_W^2) \sin (\alpha + \beta) 
+ \frac{g m_b^2 \sin \alpha}{m_W \cos \beta} &
\frac{g m_b}{2 m_W \cos \beta} (\mu \cos \alpha + A_b \sin \alpha) \\
\frac{g m_b}{2 m_W \cos \beta} (\mu \cos \alpha + A_b \sin \alpha) &
- \frac{g m_Z}{c_W}\, \frac{1}{3} \, s_W^2 \sin (\alpha + \beta) 
+ \frac{g m_b^2 \sin \alpha}{m_W \cos \beta}
\end{array}\right) \nn \\
\end{eqnarray}

For $H^o$
\begin{equation}
A_{H^o} = \left( \begin{array}{cc}
a_{H^o}^{\tilde t} & 0 \\ 0 & a_{H^o}^{\tilde b} 
\end{array}\right) 
\end{equation}

where

\begin{eqnarray}
a_{H^o}^{\tilde t} &=& \left( \begin{array}{cc}
-\frac{g m_Z}{c_W}(\frac{1}{2} - \frac{2}{3} s_W^2) \cos (\alpha + \beta) 
- \frac{g m_t^2 \sin \alpha}{m_W \sin \beta} &
-\frac{g m_t}{2 m_W \sin \beta} (- \mu \cos \alpha + A_t \sin \alpha) \\
-\frac{g m_t}{2 m_W \sin \beta} (- \mu \cos \alpha + A_t \sin \alpha) &
-\frac{g m_Z}{c_W} \frac{2}{3} s_W^2 \cos (\alpha + \beta) 
- \frac{g m_t^2 \sin \alpha}{m_W \sin \beta}
\end{array}\right) \nn \\
a_{H^o}^{\tilde b} &=& \left( \begin{array}{cc}
- \frac{g m_Z}{c_W}(- \frac{1}{2} + \frac{1}{3} s_W^2) \cos (\alpha + \beta) 
- \frac{g m_b^2 \cos \alpha}{m_W \cos \beta} &
- \frac{g m_b}{2 m_W \cos \beta} (- \mu \sin \alpha + A_b \cos \alpha) \\
- \frac{g m_b}{2 m_W \cos \beta} (- \mu \sin \alpha + A_b \cos \alpha) &
\frac{g m_Z}{c_W}\, \frac{1}{3} \, s_W^2 \cos (\alpha + \beta) 
- \frac{g m_b^2 \cos \alpha}{m_W \cos \beta}
\end{array}\right) \nn \\
\end{eqnarray}

For $A^o$
\begin{equation}
A_{A^o} = \left( \begin{array}{cc}
a_{A^o}^{\tilde t} & 0 \\ 0 & a_{A^o}^{\tilde b} 
\end{array}\right) 
\end{equation}

where

\begin{eqnarray}
a_{A^o}^{\tilde t} &=& \left( \begin{array}{cc}
0 & 
\frac{i g m_t}{2 m_W} (\mu + A_t \cot \beta) \\
- \frac{i g m_t}{2 m_W} (\mu + A_t \cot \beta) &
0
\end{array}\right) \nn \\ 
a_{A^o}^{\tilde b} &=& \left( \begin{array}{cc}
0 & 
\frac{i g m_b}{2 m_W} (\mu + A_b \tan \beta) \\
- \frac{i g m_b}{2 m_W} (\mu + A_b \tan \beta) &
0
\end{array}\right) 
\end{eqnarray}

For $H^1$ and $H^2$

\begin{equation}
A_{H^1} = \left( \begin{array}{cc}
0 & a_{H^1} \\ a_{H^1}^T & 0 
\end{array}\right) \, \, , \, \, 
A_{H^2} = \left( \begin{array}{cc}
0 & a_{H^2} \\ -a_{H^2}^T & 0 
\end{array}\right) 
\end{equation} 

where

\begin{eqnarray}
a_{H^1} &=& \left( \begin{array}{cc}
\frac{g}{2 m_W}(m_b^2 \tan\beta + m_t^2 \cot\beta - m_W^2 \sin 2\beta)
& \frac{g m_b}{2 m_W}(\mu + A_b \tan\beta) \\
\frac{g m_t}{2 m_W}(\mu + A_t \cot\beta) &
\frac{g m_t m_b}{2 m_W}(\tan\beta + \cot\beta)
\end{array}\right) \nn \\ 
a_{H^2} &=& i a_{H^1} 
\end{eqnarray}

\section*{{\bf{B}} Large mass expansion of loop integrals and mixing angles}
\renewcommand{\theequation}{B.\arabic{equation}}


Here we give the expansions of the loop integrals and mixing angles in
inverse powers of the SUSY mass parameters. 
For the two- and three-point integrals that appear in
Eqs.(\ref{eq.twopoint}) and (\ref{eq.threepoint}), we
follow the definitions and conventions of~\cite{Hollik}. The
integrals are performed in $D=4 - \epsilon$ dimensions and the divergent
contributions are regularized by $\Delta \equiv 2/\epsilon - \gamma_E +
\log(4\pi) - log(M_{SUSY}^2/\mu_0^2)$, with the corresponding
inclusion of the energy scale given here by $\mu_0$.

The two-point integrals are given by
\begin{equation}
    \mu_0^{4-D} \int \frac{d^D k}{(2\pi)^D} \frac{\{1; k^{\mu}\}}
    {[k^2 - m_1^2][(k+q)^2 - m_2^2]}
    = \frac{i}{16\pi^2} \left\{ B_0; q^{\mu} B_1 \right\}
    (q^2;m_1^2,m_2^2)\,.
\end{equation}
The three-point integrals are given by
\bea
    &\mu_0^{4-D}& \int \frac{d^D k}{(2\pi)^D}
    \frac{\{1; k^{\mu}\}}
    {[k^2 - m_1^2][(k + p_1)^2 - m_2^2][(k + p_1 + p_2)^2 - m_3^2]}
    \nonumber \\
    & & = \frac{i}{16\pi^2}
    \left\{ C_0; p_1^{\mu}C_{11} + p_2^{\mu}C_{12} \right\}
    (p_1^2, p_2^2, p^2; m_1^2, m_2^2, m_3^2)\,,
\eea

\vspace{-0.1in}\noindent
where $p = -p_1 - p_2$.

We are considering here the limit where all the squarks and gluinos
are nearly degenerate and very heavy as compared to the electroweak
scale. In terms of the MSSM parameters, this limit is reached by
choosing $M_{SUSY} \sim M_{\tilde g} \sim  M_{\tilde Q} \sim 
M_{\tilde U} \sim  M_{\tilde D} \sim \mu  \gg m_{EW}$, and the
physical squark masses can be written as 
$M_{\tilde q_{1,2}}^2 = M_{SUSY}^2 \pm m_q X_q$.
The results of the large $M_{SUSY}$ expansions of the loop integrals
in this limit are as follows
\bea
\label{eq:maxbt}
&& C_0(p_q^2,p_{H}^2,p_{q'}^2;M_{\tilde g}^2, M_{\tilde q_i}^2, 
M_{\tilde q_j}^2) \simeq -\frac{1}{2 M_{SUSY}^2} + \mathcal{O}\left(
\frac{m_{EW}}{M_{SUSY}^3} \right) \nn\\
&& C_{11}(p_q^2,p_{H}^2,p_{q'}^2;M_{\tilde g}^2, M_{\tilde q_i}^2,
    M_{\tilde q_j}^2) \simeq \frac{1}{3 M_{SUSY}^2} + \mathcal{O}\left(
\frac{m_{EW}}{M_{SUSY}^3} \right) \nn \\
&& C_{12}(p_q^2,p_{H}^2,p_{q'}^2;M_{\tilde g}^2,M_{\tilde q_i}^2,
M_{\tilde q_j}^2) \simeq \frac{1}{6 M_{SUSY}^2} + \mathcal{O}\left(
\frac{m_{EW}}{M_{SUSY}^3} \right)  \nn\\
&&B_0(p_q^2;M_{\tilde g}^2,M_{\tilde q_i}^2)
    \simeq \Delta 
+ (-1)^i \frac{m_q X_q}{2 M_{SUSY}^2} + \mathcal{O}\left(
\frac{m_{EW}^2}{M_{SUSY}^2} \right) \nn \\ 
&&B_1(p_q^2;M_{\tilde g}^2,M_{\tilde q_i}^2)
    \simeq - \frac{1}{2} \Delta  
- (-1)^i \frac{m_q X_q}{3 M_{SUSY}^2}  + \mathcal{O}\left(
\frac{m_{EW}^2}{M_{SUSY}^2} \right), \nn \\ 
\label{eq.expintegrals}
\eea
where $X_q$ are defined in Eq.~(\ref{eq.MLRtb}), $p_q,p_{q'}$ and
$p_{H}$ are the external momenta of quarks and Higgs particles,
respectively, and $m_{EW}$ refers generically to any external momentum
or mass of the order of the electroweak scale.

The expansions for the mixing angles in the large $M_{SUSY}$ limit, are as follows:
\begin{eqnarray}
\sin 2\theta_q \simeq 1 + \mathcal{O}\left( 
\frac{m_{EW}^2}{M_{SUSY}^2}\right) \, \, &,& \, \,
\cos 2\theta_q \simeq 0 + \mathcal{O}\left( 
\frac{m_{EW}}{M_{SUSY}} 
\right),
\label{eq.expangles}
\end{eqnarray}
thus, our choice of large MSSM parameters above implies nearly
maximal squark mixing. 

Finally, notice that we have written explicitly in
Eqs.(\ref{eq.expintegrals}) and (\ref{eq.expangles}) only the leading
terms in the expansions. The next to
leading terms of these expansions can be found in~\cite{HaberTemes,ourHtb}.
In particular, because of our choice of $\mu \sim M_{SUSY}$, $X_q$ are of
$\mathcal{O}(M_{SUSY})$ and, therefore, the second terms in the
expansions of $B_0$ and $B_1$ are of $\mathcal{O}(m_{EW}/M_{SUSY})$.
For the $B_0$ case, this term will also contribute, via
Eq.(\ref{eq.twopoint}) to the non-decoupling effects in the quark
propagators. Regarding the three-point functions, it is $C_0$ that via
Eq.(\ref{eq.threepoint}) will contribute to the non-decoupling terms.

\begingroup\raggedright\endgroup

\end{document}